\documentclass[letter, longauth]{aa}
\usepackage[varg]{txfonts}
\usepackage{natbib}
\usepackage{graphicx}
\usepackage{amsmath}
\usepackage{amssymb}
\usepackage{ragged2e}
\usepackage{placeins}
\usepackage[
    separate-uncertainty=true,
    multi-part-units=single,
    range-phrase=\mbox{--},
    range-units=single
]{siunitx}
\usepackage{hyperref}
\usepackage{xfrac}

\usepackage{lscape}

\usepackage{multirow}
\usepackage{array}
\newcommand{\PreserveBackslash}[1]{\let\temp=\\#1\let\\=\temp}
\newcolumntype{C}[1]{>{\PreserveBackslash\centering}p{#1}}

\DeclareSIUnit{\arcsec}{arcsec}
\DeclareSIUnit{\mas}{mas}
\DeclareSIUnit{\au}{au}
\DeclareSIUnit{\year}{yr}
\DeclareSIUnit{\solarmass}{M\textsubscript{\(\odot\)}}
\DeclareSIUnit{\solarradius}{R\textsubscript{\(\odot\)}}
\DeclareSIUnit{\earthmass}{M\textsubscript{\(\oplus\)}}
\DeclareSIUnit{\earthradius}{R\textsubscript{\(\oplus\)}}
\DeclareSIUnit{\ppt}{ppt}
\DeclareSIUnit{\btjd}{BTJD}
\DeclareSIUnit{\electron}{e\textsuperscript{\(-\)}}

\bibpunct{(}{)}{;}{a}{}{,}

\defcitealias{barnard}{GH24} 
\begin{document}
\title{A sub-Earth-mass planet orbiting Barnard's star}
\subtitle{No evidence of transits in TESS photometry}

\titlerunning{No evidence of transits around Barnard's star}
\authorrunning{Stefanov et al.}

\author{
A. K. Stefanov\inst{1,2,*},
J. I. Gonz\'alez Hern\'andez\inst{1,2},
A. Su\'arez Mascare\~no\inst{1,2},
N. Nari\inst{1,2,3},
R. Rebolo\inst{1,2,4},
M. Damasso\inst{5},
A. Castro-Gonz\'alez\inst{6},
M.-R. Zapatero Osorio\inst{6},
C. Allende Prieto\inst{1,2},
A. M. Silva\inst{7,8,9},
C. J. A. P. Martins\inst{7,10}
}
\date{Accepted 28 November 2024}

\institute{
\inst{1}Instituto de Astrof\'isica de Canarias, 38205 La Laguna, Spain\\
\inst{2}Departamento de Astrof\'isica, Universidad de La Laguna, 38206 La Laguna, Spain\\
\inst{3}Light Bridges S. L., 35004 Las Palmas de Gran Canaria, Spain\\
\inst{4}Consejo Superior de Investigaciones Cient\'ificas (CSIC), 28006 Madrid, Spain\\
\inst{5}INAF - Osservatorio Astrofisico di Torino, Via Osservatorio 20, 10025 Pino Torinese, Italy\\
\inst{6}Centro de Astrobiolog\'ia (CAB), CSIC-INTA, ESAC campus, Camino Bajo del Castillo s/n, 28692 Villanueva de la Ca\~nada, Spain\\
\inst{7}Instituto de Astrof\'isica e Ci\^encias do Espa\c{c}o, CAUP, Universidade do Porto, Rua das Estrelas, 4150-762 Porto, Portugal\\
\inst{8}Departamento de F\'isica e Astronomia, Faculdade de Ci\^encias, Universidade do Porto, Rua do Campo Alegre, 4169-007 Porto, Portugal\\
\inst{9}Instituto de Astrof\'isica e Ci\^encias do Espa\c{c}o, Faculdade de Ci\^encias da Universidade de Lisboa, Campo Grande, 1749-016 Lisboa, Portugal\\
\inst{10}Centro de Astrof\'isica da Universidade do Porto, Rua das Estrelas, 4150-762 Porto, Portugal\\
\inst{*}\email{atanas.stefanov@iac.es}
}

\abstract{A sub-Earth-mass planet orbiting Barnard's star, designated as Barnard~b, has  recently been announced. At almost the same time, the first photometric data of Barnard's star by the Transit Exoplanet Survey Satellite (TESS) was released in Sector 80. We explore the possibility of emergent transits of Barnard~b in TESS photometry. The detrended 2 min light curve appears to be flat, with a flux root mean square of 0.411 parts per thousand. Attempts of blind and informed transit curve model inference suggest no evidence of transiting Barnard~b, or any other body. This provides a 3$\sigma$ upper bound of 87.9 degrees for the orbital inclination of Barnard~b.}


\maketitle

\section{Introduction}

Exoplanetary astronomy has had a remarkable expansion in recent years. The number of confirmed exoplanets will soon reach \num{5800} (NASA Exoplanet Archive, \citealp{nasaexoplanettable}; accessed November 2024). Astronomers are already on track to discover potentially habitable Earth-like planets (e.g. \citealp{Crossfield2015,Gillon2016,Murgas2023,SuarezMascareno2023}). The discovery of a planet in the habitable zone of Proxima, in the closest stellar system to our own (Proxima~b; \citealp{Anglada-Escude2016}), rocked the scientific community. It hinted that potential Earth analogues may lie in the very vicinity of the solar neighbourhood. Subsequently, considerable effort has been devoted to the analysis of these nearby systems. This led to a natural focus on the second-closest stellar system: Barnard's star (Table~\ref{tab:stellar_parameters}), an M3.5V-M4V dwarf that has been thoroughly examined since its discovery \citep{Barnard1916}.

More recently, \citet{barnard}, hereafter \citetalias{barnard}, reported the detection of a sub-Earth-mass planet around Barnard's star. This planet, designated as Barnard b, has an orbital period $P_\text{b}=\SI{3.1533(0.0006)}{\day}$ and a minimum mass of $M_\text{b}\sin i_\text{b} = \SI{0.37(0.05)}{\earthmass}$. The proximity of Barnard~b to its host star (\SI{0.02}{\au}) implies a transit probability of 3.7\%,  which in turn would offer an extraordinary opportunity for its atmospheric characterisation. The Transiting Exoplanet Survey Satellite (TESS; \citealp{tess}) had obtained photometric observations of Barnard's star between June and July 2024, and released those data as part of the Sector 80 (S80) data release in late August 2024,  a mere few days after the acceptance of \citetalias{barnard}. The present Letter serves to extend their work by exploring potential transits of Barnard~b in this new and only TESS photometry.

\begin{table}
\centering
\caption{
Relevant parameters of Barnard's star.
}
\label{tab:stellar_parameters}
\begin{tabular}{lcclc}
\hline\hline
\multicolumn{1}{c}{Parameter} &
\multicolumn{1}{c}{Symbol} &
\multicolumn{1}{c}{Unit} &
\multicolumn{1}{c}{Value} &
\multicolumn{1}{c}{Ref.} \\ \hline
Right ascension &
$\alpha_\text{J2000}$ &
-- &
\phantom{+}17\textsuperscript{h}57\textsuperscript{m}48.5\textsuperscript{s} &
2 \\
Declination &
$\delta_\text{J2000}$ &
-- &
+04\degr41'36.1'' &
2 \\
Eff. temperature &
$T_\text{eff}$ &
K &
\num{3195(28)} &
1 \\
Mass &
$M_\star$ &
\unit{\solarmass} &
\num{0.162(7)} &
3 \\
Radius &
$R_\star$ &
\unit{\solarradius} &
\num{0.185(6)}&
3 \\
Surface gravity &
$\log_{10}g$ &
cm$\,$s$^{-2}$ &
\num{4.90(0.09)}&
1 \\
Metallicity &
[Fe/H] &
dex &
\num{-0.56(0.07)}&
1 \\
\hline
\end{tabular}
\tablebib{
(1) \citet{barnard};
(2) \citet{gaiaDR3};
(3) \citet{Schweitzer2019}.
}
\end{table} 
\section{Data}\label{sec:data}
The TESS Science Processing Operations Center pipeline (SPOC; \citealp{spoc}) provides two different photometric data products: Simple Aperture Photometry (SAP) and Pre-Search Data Conditioning Simple Aperture Photometry (PDCSAP). SAP is a simple summation of flux from the optimal aperture of the target determined by SPOC. Its data may suffer from systematics of different kinds, including contamination from other sources, cosmic rays, variable incident solar radiation, foreign particles on CCD sensors, and systematics from spacecraft motion. With a similar rationale, but regarding the Kepler mission, \citet{Kinemuchi2012} argued that SAP data should be expected to carry all these systematics; analysis of subtle photometric variations on timescales larger than \SI{1}{\day} is likely to be affected. PDCSAP, on the other hand, aims to mitigate signal artefacts and instrumental effects in SAP data. The basic mode of operation is described in Sect.~4.2 of the same work, and has been further refined since then (see \citealp{Stumpe2012,Smith2012,Stumpe2014}). We discuss \mbox{SAP} and \mbox{PDCSAP} data in all the analysis that follows.

We studied the origin of flux inside the SPOC aperture of Barnard's~star through the TESS contamination tool \textsc{TESS-cont} \citep{Castro-Gonzalez2024}. We analysed the TESS pixel-response functions of all sources within \SI{500}{\arcsec} \mbox{($N=4819$)}. We found that 99.5\% of the aperture flux indeed comes from Barnard's~star. The remaining 0.5\% are ascribed to fainter sources, whose small contributions add up to a negligible contamination.

\section{Analysis}\label{sec:analysis}
Two parameters define the general shape of a transit: the transit width $W$ from first to fourth contact, and the transit depth $D$. Deriving some characteristic values of $W$ and $D$, even if approximate, would give us clues when we look for features in photometry. Both parameters, but especially $D$, are sensitive to the planetary radius $R_\text{b}$, which in the case of Barnard~b is not directly probed in \citetalias{barnard}. Nevertheless, we can relate their measured $M_\text{b}\sin i_\text{b}$ to $R_\text{b}$ for an assumed mass-radius relation and an assumed impact-parameter distribution. We used the mass-radius relation provided by \textsc{spright} \citep{spright} to estimate $R_\text{b}$ under the assumption $\sin i_\text{b}\approx 1$. For an impact-parameter distribution, we adopted \mbox{$b=\mathcal{U}(0,1)$}, which would be implied by a random orbital orientation relative to the line of sight.

We then set out to estimate $W$ and $D$ through rapid transit curve modelling. We assumed that the limb-darkening profile of Barnard's star follows the quadratic law
\begin{equation}
    \frac{I(\vartheta)}{I(\vartheta=\SI{0}{\degree})}=1-u_1(1-\cos\vartheta)-u_2(1-\cos\vartheta)^2,
\end{equation}
where $I$ is the total emergent intensity, and $\vartheta$ is the angle between the line of sight and the direction of the emergent radiation \citep{Kopal1950}. We fed stellar-parameter measurements from Table~\ref{tab:stellar_parameters} to \textsc{ldtk} \citep{ldtk1,ldtk2}, which returned \mbox{$u_1=\num{0.2470(0.0013)}$} and
\mbox{$u_2=\num{0.3814(0.0026)}$} for the TESS passband. We then computed \num{e6} transit curves from the distributions of 
(i)~$R_\star, M_\star, P_\text{b},M_\text{b}$ from \citetalias{barnard};
(ii)~$R_\text{b}$ from \textsc{spright}; (iii) $u_1, u_2$ from \textsc{ldtk}, assumed Gaussian; and 
(iv)~\mbox{$b=\mathcal{U}(0,1)$}.
We calculated $W$ and $D$ of each light curve individually; we list their aggregate statistics together with our derived $R_\text{b}$ in Table~\ref{tab:planetary_parameters}. The bulk of the distribution of $R_\text{b}$ appeared Gaussian-like, with few outliers beyond \SI{1}{\earthradius}. We expect a transit of depth \mbox{$1.55^{+0.26}_{-0.28}\,$\unit\ppt} (parts per thousand) and of width
\mbox{$0.82^{+0.11}_{-0.27}\,$\unit\hour}.
\citetalias{barnard} provided with an additional four-planet solution that contains Barnard~b and three planetary candidates. We repeated the same procedure for those solutions; we list the corresponding results in Table~\ref{tab:multiplanet_parameters}.

\begin{table}
\centering
\caption{
Relevant parameters of Barnard b.
}
\label{tab:planetary_parameters}
\begin{tabular}{lclc}
\hline\hline
\multicolumn{1}{c}{Symbol} &
\multicolumn{1}{c}{Unit} &
\multicolumn{1}{c}{Value} &
\multicolumn{1}{c}{Ref.} \\ \hline
$a_\text{b}$ &
au &
0.02294(33) &
1 \\
$P_\text{b}$ &
d &
3.1533(6) &
1 \\
$T_\text{0, b}-\num{2460000}$ &
d &
\num{139.204(149)} &
1 \\
$M_\text{b}\sin i_\text{b}$ &
\unit{\earthmass} &
\num{0.37(5)} &
1 \\
$R_\text{b}$ &
\unit{\earthradius} &
\num{0.76(4)} &
0 \\
$D_\text{b}$ &
ppt &
$1.55^{+0.26}_{-0.28}$ &
0 \\
$W_\text{b}$ &
h &
$0.82^{+0.11}_{-0.27}$ &
0 \\
\hline
\end{tabular}
\tablebib{
(0) This work;
(1) \citet{barnard}.
}
\tablefoot{
The expected radius, transit depth, and transit width were naïvely computed under two assumptions:
(i)~that Barnard~b follows the mass-radius relation provided by \citet{spright} and 
(ii)~that the impact parameter follows the uniform distribution $\mathcal{U}(0,1)$. Reported uncertainties reflect the 16{th} and the 84{th} percentiles. Some uncertainties are given in parentheses, to the order of the least significant figure.
}
\end{table} 
\subsection{Lack of features in TESS S80 data}

\begin{figure}
    \centering
    \resizebox{\hsize}{!}{\includegraphics{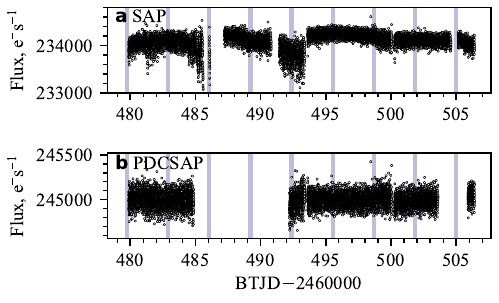}}
    \caption{TESS S80 photometric data products: (a)~SAP; (b)~\mbox{PDCSAP}, with superimposed predicted times of inferior conjunction through \citetalias{barnard} ephemerides. The propagated inferior-conjunction uncertainty is shown as light blue bands.}
    \label{fig:timeseries}
\end{figure}

We took all \SI{2}{\minute} SAP and PDCSAP measurements from the TESS S80 lightcurve, using the default cadence quality bitmask. The SAP dataset is comprised of \num{16492} measurements in the range \SIrange{2460479}{2460507}{\btjd}. SAP measurements have a root mean square (rms) of \SI{166}{\electron\per\second} and a median uncertainty of \SI{52}{\electron\per\second}. We show these time series in Fig.~\ref{fig:timeseries}a, together with predicted inferior conjunctions of Barnard~b from \citetalias{barnard} emphemerides. 
 SAP measurements are split into several portions, and contain irregular negative trends that tend to take place at the end of those portions (e.g. near \SI{2460485}{\btjd} or \SI{2460493}{\btjd}). However, there is no apparent evidence of a strictly periodic decrease in flux, which would otherwise be caused by the sought planetary transit. The PDCSAP time series share the same temporal coverage, but with \num{4691} excluded measurements in the interval \SIrange{2460485}{2460492}{\btjd} (Fig.~\ref{fig:timeseries}b). This exclusion affected the coverage of two predicted inferior conjunctions. PDCSAP time series contain \num{11801} measurements that have a flux rms of \SI{78.4}{\electron\per\second} and a median uncertainty of \SI{54.5}{\electron\per\second}.

We proceeded with SAP in an attempt to include data near the inferior-conjunction predictions that PDCSAP dismissed. We smoothed the SAP time series with several variants of detrending, including second-order Savitzky-Golay filters \citep{savgol}, Tukey's biweight time-window sliders \citep{Mosteller1977}, and mean time-window sliders,  all available in \textsc{Wōtan} \citep{wotan}. We globally used a window length of \SI{0.5}{\day}, an edge cut-off of \SI{0.3}{\day}, and a break tolerance of \SI{0.2}{\day}. For none of these forms of detrending could we find evidence of transits after generating transit least squares periodograms \mbox{(TLSPs;} \citealp{tls}) or box least squares periodograms \mbox{(BLSPs;} \citealp{bls}) for periodicities in the range \SIrange{1}{15}{\day} and transit widths in the range \SIrange{0.2}{2}{\hour}.

\begin{figure*}
    \centering
    \resizebox{\hsize}{!}{\includegraphics{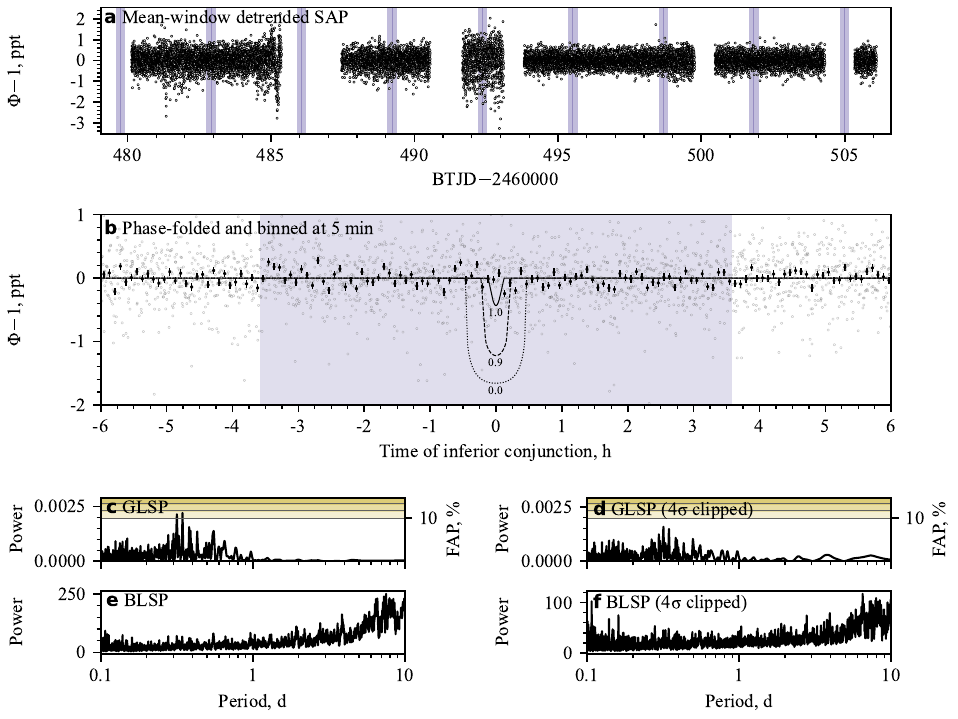}}
    \caption{
    The search of Barnard~b in TESS S80 photometry. (a)~Mean time-window detrended SAP photometry with superimposed predicted times of inferior conjunction through \citetalias{barnard} ephemerides. The median expected transit width is displayed as dark blue bands. The propagated ephemeris uncertainty is shown as light blue bands.
    (b)~Phase-folded plot at \citetalias{barnard} ephemerides: before and after applying a \SI{5}{\minute} binning (light hollow and dark filled circles, respectively). The inferior-conjunction epoch uncertainty is shown as a light blue band. We compare photometry against the expected transit curve of Barnard~b for three impact-parameter values: $b=0.0$ (black dotted line), $b=0.9$ (black dashed line), and $b=1.0$ (black solid line).
    (c,d)~GLSP of unbinned data before and after iterative $4\sigma$ clipping.
    (e,f)~BLSP of unbinned data before and after iterative $4\sigma$ clipping.
    }
    \label{fig:timeseries2}
\end{figure*}

We chose to go forward with mean time-window detrending by virtue of its simplicity. Figure~\ref{fig:timeseries2} displays this detrended SAP photometry.
 It contains \num{14366} measurements that are characterised by a flux rms of \SI{0.411}{\ppt} and a median uncertainty of \SI{0.222}{\ppt}. The flux rms is nearly four times smaller than the expected transit depth
\mbox{($1.55^{+0.26}_{-0.28}$\,\unit\ppt;}
Fig.~\ref{tab:planetary_parameters}). Figure~\ref{fig:timeseries2}a compares these processed time series again with the predicted inferior conjunctions of Barnard~b from \citetalias{barnard}. Six inferior conjunctions are covered by photometry within the  ephemeris precision. We could observe no significant decreases in flux at those ephemerides or anywhere else. We folded the time series at the published period and epoch of Barnard~b, and binned the folded data at \SI{5}{\minute}. Figure~\ref{fig:timeseries2}b illustrates the result of this exercise, together with the epoch uncertainty and \textsc{pytransit} curves from the median posteriors of Table~\ref{tab:planetary_parameters} for three impact-parameter values: 0, 0.9, and 1. The measurements clearly do not  follow predictions for such a liberal range of impact parameters, regardless of whether binning was involved or not. This remained the case after binning at other intervals (\SIrange{2}{10}{\minute}). On the whole, these timeseries give an impression of notable symmetry and lack of apparent features, as if we  were observing pure noise when measuring a constant-flux source. The distribution of measurements is Gaussian-like, with its right tail being asymmetrically heavier (Fig.~\ref{fig:pdf}). On the other hand, transits must break the flux distribution symmetry in the opposite direction, i.e. in the left tail.

We looked for any out-of-transit variations through the generalised Lomb-Scargle periodogram (GLSP; \citealp{Lomb1976,Scargle1982,Zechmeister2009,VanderPlas2015}) of the detrended SAP photometry. Figure~\ref{fig:timeseries2}c displays this exercise. The GLSP highlights two peaks near \SI{0.32}{\day} and \SI{0.35}{\day}, and assigns them a false-alarm probability (FAP; \citealp{Baluev2008}) of 3\% and 2\%, respectively. Their periodicities do not appear to be related to the window length we used (\SI{0.5}{\day}). Phase-folding at those periods did not reveal any apparent structures in photometry. Analysis in the spirit of \citet{Dawson2010} revealed that the common harmonics of the two peaks do not correspond to any significant peaks in the window function of the dataset, and therefore are unlikely to be aliases of one another. We found out that the iterative sigma-clipping of the SAP photometry suppressed these signals greatly (Figure~\ref{fig:timeseries2}d). Clipping within
$5\sigma$, $4\sigma$, and $3\sigma$ resulted in GLSPs where the strongest peak had a FAP of 5\%, 78\%, and >99\%, respectively. Most of the rejected outliers tend to congregate around the first (\SIrange{2460480}{2460485}{\btjd}) and third seasons (\SIrange{2460492}{2460493}{\btjd}) in regions of higher flux dispersion. All of these observations, together with the demonstrated good behaviour of the flux distribution, are against the physical nature of the \SI{0.32}{\day} and \SI{0.35}{\day} signals. Figures~\ref{fig:timeseries2}e and \ref{fig:timeseries2}f show the BLSPs of the detrended SAP photometry before and after sigma-clipping. In neither periodogram could we find any significant peaks that merit discussion.

\subsection{Lack of plausible transit solutions}

We performed transit-curve fitting through \textsc{pytransit} and nested sampling \citep{Skilling2004}.
For the inference of these models, we used the nested-sampling integrator provided by \textsc{Dynesty} \citep{dynesty}. In every inference instance of $N_\text{param}$ model parameters, we required $40N_\text{param}$ live points and random-walk sampling. We modelled one planet on a circular orbit, using \citetalias{barnard} ephemerides as priors. We imposed two different priors on the planetary radius: one informed by the expected radius
\mbox{$\mathcal{N}(0.76,0.04)\,\unit\earthradius$} from Table~\ref{tab:planetary_parameters}, and one with the very uninformed $\mathcal{U}_\text{log}(e^{-5},e^{+5})\,\unit\earthradius$ (i.e. up to \SI{148}{\earthradius}). We hereafter refer to these priors as the narrow and wide priors, respectively. Our impact-parameter prior was liberally set to $\mathcal{U}(0,1.5)$. Quadratic limb-darkening coefficients took priors defined by the Gaussian distributions defined by \textsc{ldtk}, given the stellar parameters from Table~\ref{tab:stellar_parameters}. Their variances were scaled three-fold in the prior to ensure impartiality. All models sampled for a zero-order flux offset $\Delta\Phi$, which pinpoints the  out-of-transit flux level relative to $\Phi=1$. Additionally, all models included a rigid jitter component of \SI{0.25}{\ppt}, based on the rms of the first \SI{10}{\day} of data.

Bayesian evidence of one-planet models reveals that the data do not support a planetary addition
(\mbox{$\Delta\ln Z=-89$} for narrow $R_\text{b}$ priors,
\mbox{$\Delta\ln Z=-143$} for wide $R_\text{b}$ priors). The narrow-prior model restricts impact-parameter solutions to $b>1$ $(3\sigma)$, which correspond to inclinations $i_\text{b}<\SI{87.9}{\degr}$ (Figure~\ref{fig:posterior}). This suggests that the data are not compatible with any transiting events with the properties of Barnard~b, including cases of grazing transits. Although the wide-prior model provided upper limits to potential transiting planets, we note a large degeneracy between the  radius and the inclination. For $i_\text{b}>\SI{88}{\degr}$, planetary radii up to a few tenths of the $\unit\earthradius$ were allowed. For less aligned transits, sensitivity decreased to the order of $\unit\earthradius$,  to such an extent that even \SI{10}{\earthradius} planets would remain undetected for $i_\text{b}<\SI{87}{\degr}$  (Fig.~\ref{fig:detection_lims}). We tested models with and without Gaussian processes (GPs; \citealp{Rasmussen2006}), using a Matérn 5/2 kernel, as defined in \textsc{s+leaf} \citep{spleaf1, spleaf2}. We additionally included three more planets in the model to try to reach the four-planet solution in \citetalias{barnard}. The inclusion of neither GPs nor additional planets had a meaningful effect.

\section{Conclusion}\label{sec:conclusion}
We investigated the first and, to date, only TESS sector, S80, that contains photometry of Barnard's star. PDCSAP and SAP photometry were shown to carry no apparent transit features within measurement precision. The flux distribution in detrended SAP photometry appears Gaussian-like.
We attempted transit-curve fitting with informed and uninformed priors on the planetary radius, but this yielded no significant results. We applied a very similar methodology on detrended PDCSAP photometry of \SI{20}{\second} and of \SI{2}{\minute} cadence;  both proved equally unfruitful. On that account, we conclude that current TESS photometry strongly suggests that Barnard~b is non-transiting.
This implies a marginal constraint in orbital inclination of the planet ($i_\text{b}\leq 87.9\degr$; $3\sigma$). We were unable to find any other periodic darkening in the photometry time series that could be ascribed to planetary transits.

\begin{acknowledgements}
AKS acknowledges the support of a fellowship from the ``la Caixa'' Foundation (ID 100010434). The fellowship code is LCF/BQ/DI23/11990071.
AKS, JIGH, ASM, CAP, NN, and RR acknowledge financial support from the Spanish Ministry of Science and Innovation (MICINN) project PID2020-117493GB-I00 and from the Government of the Canary Islands project ProID2020010129.
NN acknowledges funding from Light Bridges for the Doctoral Thesis ``Habitable Earth-like planets with ESPRESSO and NIRPS'',
in cooperation with the Instituto de Astrofísica de Canarias, and the use of Indefeasible Computer Rights (ICR) being commissioned at the ASTRO POC project in Tenerife, Canary Islands, Spain. The ICR-ASTRONOMY used for his research was provided by Light Bridges in cooperation with Hewlett Packard Enterprise (HPE).
ACG is funded by the Spanish Ministry of Science through MCIN/AEI/10.13039/501100011033 grant PID2019-107061GB-C61. 
AMS acknowledges support from the Fundação para a Ciência e a Tecnologia (FCT) through the Fellowship 2020.05387.BD (DOI: 10.54499/2020.05387.BD). Funded/Co-funded by the European Union (ERC, FIERCE, 101052347). Views and opinions expressed are however those of the author(s) only and do not necessarily reflect those of the European Union or the European Research Council. Neither the European Union nor the granting authority can be held responsible for them. This work was supported by FCT - Fundação para a Ciência e a Tecnologia through national funds by these grants: UIDB/04434/2020, UIDP/04434/2020.
The work of CJAPM was financed by Portuguese funds through FCT (Funda\c c\~ao para a Ci\^encia e a Tecnologia) in the framework of the project 2022.04048.PTDC (Phi in the Sky, DOI 10.54499/2022.04048.PTDC). CJAPM also acknowledges FCT and POCH/FSE (EC) support through Investigador FCT Contract 2021.01214.CEECIND/CP1658/CT0001 (DOI 10.54499/2021.01214.CEECIND/CP1658/CT0001). 

This paper includes data collected by the TESS mission, which are publicly available from the Mikulski Archive for Space Telescopes (MAST). Funding for the TESS mission is provided by NASA's Science Mission Directorate. Resources supporting this work were provided by the NASA High-End Computing (HEC) Program through the NASA Advanced Supercomputing (NAS) Division at Ames Research Center for the production of the SPOC data products. We acknowledge the use of public TESS data from pipelines at the TESS Science Office and at the TESS Science Processing Operations Center. This research has made use of the Exoplanet Follow-up Observation Program website, which is operated by the California Institute of Technology, under contract with the National Aeronautics and Space Administration under the Exoplanet Exploration Program.
This work made use of \textsc{TESS-cont}, which builds upon \textsc{tpfplotter} \citep{Aller2020} and \textsc{TESS\_PRF} \citep{Bell2022}.
We used the following \textsc{Python} packages for data analysis and
visualisation:
\textsc{Astropy} \citep{astropy},
\textsc{dynesty} \citep{dynesty},
\textsc{Lightkurve} \citep{lightkurve},
\textsc{ldtk} \citep{ldtk1,ldtk2},
\textsc{Matplotlib} \citep{matplotlib},
\textsc{nieva} (Stefanov et al., in prep.),
\textsc{NumPy} \citep{numpy},
\textsc{pandas} \citep{pandas1,pandas2},
\textsc{pyastronomy} \citep{pyastronomy},
\textsc{pytransit} \citep{pytransit},
\textsc{SciPy} \citep{scipy},
\textsc{seaborn} \citep{seaborn},
\textsc{s+leaf} \citep{spleaf1,spleaf2},
\textsc{spright} \citep{spright} and
\textsc{Wōtan} \citep{wotan}.
This manuscript was written and compiled in \textsc{Overleaf}.\end{acknowledgements}

\bibliographystyle{aa}
\bibliography{zotero_bibtex}

\begin{appendix}
\section{Supplementary material}
\begin{figure}[h]
    \centering
    \resizebox{\hsize}{!}{\includegraphics{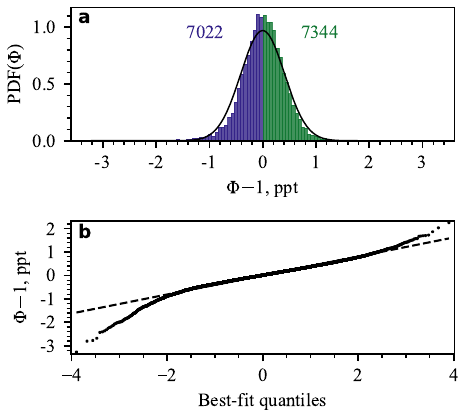}}
    \caption{Characteristics of detrended PDCSAP photometry. (a) Distribution plot, conventionally split in two into a lower distribution (blue bars) and an upper distribution (green bars). The size of these sub-distributions are given in figures of corresponding colour. A Gaussian-fuction fit to the distribution is given as a black solid line. (b) Probability plot assessing whether the full distribution (black solid line) follows the fitted Gaussian (black dashed line).}
    \label{fig:pdf}
\end{figure} \FloatBarrier
\begin{table}
\centering
\caption{
Relevant parameters of Barnard~b and planetary candidates from the four-planet solution in \citet{barnard}.
}
\label{tab:multiplanet_parameters}
\begin{tabular}{lclc}
\hline\hline
\multicolumn{1}{c}{Symbol} &
\multicolumn{1}{c}{Unit} &
\multicolumn{1}{c}{Value} &
\multicolumn{1}{c}{Ref.} \\ \hline
\textbf{Barnard b} \\
$a_\text{b}$ &
\unit{\au} &
0.0229(3) &
1 \\
$P_\text{b}$ &
\unit{\day} &
3.1537(5) &
1 \\
$T_\text{0, b}-\num{2460000}$ &
\unit{\day} &
\num{139.290(104)} &
1 \\
$M_\text{b}\sin i_\text{b}$ &
\unit{\earthmass} &
\num{0.32(4)} &
1 \\
$R_\text{b}$ &
\unit{\earthradius} &
\num{0.73(4)} &
0 \\
$D_\text{b}$ &
\unit{\ppt} &
$1.42^{+0.23}_{-0.26}$ &
0 \\
$W_\text{b}$ &
\unit{\hour} &
$0.82^{+0.11}_{-0.27}$ &
0 \\
\textbf{Barnard c (cand.)} \\
$a_\text{c}$ &
\unit{\au} &
0.0274(4) &
1 \\
$P_\text{c}$ &
\unit{\day} &
4.1243(8) &
1 \\
$T_\text{0, c}-\num{2460000}$ &
\unit{\day} &
\num{139.730(129)} &
1 \\
$M_\text{c}\sin i_\text{c}$ &
\unit{\earthmass} &
\num{0.31(4)} &
1 \\
$R_\text{c}$ &
\unit{\earthradius} &
\num{0.72(4)} &
0 \\
$D_\text{c}$ &
\unit{\ppt} &
$1.40^{+0.23}_{-0.25}$ &
0 \\
$W_\text{c}$ &
\unit{\hour} &
$0.89^{+0.12}_{-0.30}$ &
0 \\
\textbf{Barnard d (cand.)} \\
$a_\text{d}$ &
\unit{\au} &
0.0188(3) &
1 \\
$P_\text{d}$ &
\unit{\day} &
2.3407(4) &
1 \\
$T_\text{0, d}-\num{2460000}$ &
\unit{\day} &
\num{138.458(101)} &
1 \\
$M_\text{d}\sin i_\text{d}$ &
\unit{\earthmass} &
\num{0.22(3)} &
1 \\
$R_\text{d}$ &
\unit{\earthradius} &
\num{0.65(4)} &
0 \\
$D_\text{d}$ &
\unit{\ppt} &
$1.12^{+0.20}_{-0.21}$ &
0 \\
$W_\text{d}$ &
\unit{\hour} &
$0.74^{+0.10}_{-0.25}$ &
0 \\
\textbf{Barnard e (cand.)} \\
$a_\text{e}$ &
\unit{\au} &
0.0381(6) &
1 \\
$P_\text{e}$ &
\unit{\day} &
6.7377(56) &
1 \\
$T_\text{0, e}-\num{2460000}$ &
\unit{\day} &
\num{137.405(551)} &
1 \\
$M_\text{e}\sin i_\text{e}$ &
\unit{\earthmass} &
\num{0.17(5)} &
1 \\
$R_\text{e}$ &
\unit{\earthradius} &
$0.62^{+0.08}_{-0.05}$ &
0 \\
$D_\text{e}$ &
\unit{\ppt} &
$1.01^{+0.31}_{-0.21}$ &
0 \\
$W_\text{e}$ &
\unit{\hour} &
$1.05^{+0.14}_{-0.35}$ &
0 \\
\hline
\end{tabular}
\tablebib{
(0) This work;
(1) \citet{barnard}.
}
\tablefoot{
The expected radius, transit depth, and transit width were naïvely computed under two assumptions:
(i)~that Barnard~b follows the mass-radius relation provided by \citet{spright} and 
(ii)~that the impact parameter follows the uniform distribution $\mathcal{U}(0,1)$. The reported uncertainties reflect the 16{th} and the 84{th} percentiles. Some uncertainties are given in parentheses, to the order of the least significant figure.
}
\end{table} \begin{figure*}
    \centering
    \resizebox{\hsize}{!}{\includegraphics{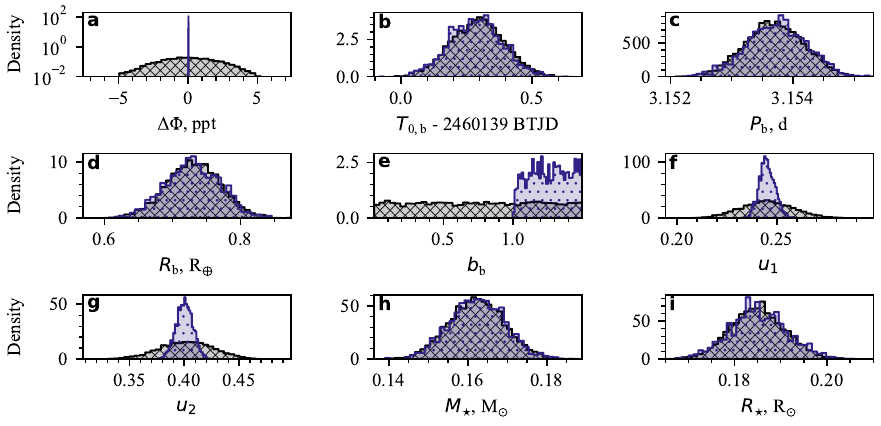}}
    \caption{Prior (black, cross-hatched) and posterior (blue, dot-hatched) parameter distributions in our narrow-prior radius search of Barnard~b.
    }
    \label{fig:posterior}
\end{figure*} \begin{figure}
    \centering
    \resizebox{\hsize}{!}{\includegraphics{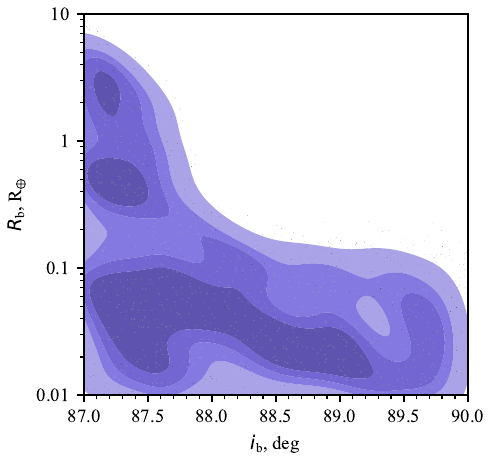}}
    \caption{Orbital-inclination and planetary-radius distributions in our wide-prior radius search of Barnard~b. The black dots mark the location of all samples in our posterior distribution. The shaded contours highlight 20\% 40\%, 60\%, and 80\% of the enclosed probability mass.}
    \label{fig:detection_lims}
\end{figure} \end{appendix}

\end{document}